\documentstyle[prl,twocolumn,epsf,floats,aps]{revtex}
\begin{document}
\draft

\twocolumn[\hsize\textwidth\columnwidth\hsize\csname
@twocolumnfalse\endcsname

\title{Spin textures, screening and excitations in dirty 
quantum Hall ferromagnets}
\author{S. Rapsch$^1$, J. T. Chalker$^1$, and D. K. K. Lee$^2$}
\address{$^1$Theoretical Physics, Oxford University, Oxford
OX1 3NP, United Kingdom\\ $^2$Blackett Laboratory, Imperial College,
Prince Consort Road, London, SW7 2BW, United Kingdom}
\date{\today}
\maketitle

\begin{abstract}
We study quantum Hall ferromagnets in the presence of a random electrostatic 
impurity potential. Describing these systems with a classical non-linear
sigma model and using analytical estimates supported by results
from numerical simulations, we examine the nature of the ground state
as a function of disorder strength, $\Delta$, 
and deviation, $\delta \nu$, of the average Landau level filling factor
from unity. Screening of an impurity potential requires distortions
of the spin configuration, and in the absence of Zeeman coupling
there is a disorder-driven, zero-temperature phase transition from a 
ferromagnet at small $\Delta$ and $|\delta \nu|$ to a spin glass at larger
$\Delta$ or $|\delta \nu|$.
We examine ground-state response functions and excitations.

\end{abstract}

 ]

Quantum Hall ferromagnets (QHFMs) are interesting especially as systems in which 
spin configurations and charge density are closely linked \cite{review}. 
At small
Zeeman energy and for Landau level filling factor $\nu$
close to unity,
this link has the celebrated consequence that the charged quasiparticles
with the lowest energy are not single electrons but skyrmions \cite{sondhi}.
These bound states of a minority-spin electron with one or many spin waves
may be viewed classically as topological excitations of an ordered
ferromagnet: in this description, the deviation of local charge density from
that of a filled and ferromagnetically polarised Landau level is proportional
to the topological density \cite{rajaraman} of the spin configuration.
For a clean QHFM with sufficiently small Zeeman energy, 
skyrmions or antiskyrmions are introduced at zero temperature on varying the
average filling factor from $\nu=1$ 
to larger or smaller values and, at
non-zero temperature, are generated thermally in pairs together with spin
waves. For dirty QHFMs,
coupling of an electrostatic impurity potential
to the charge density offers an additional mechanism by which spin textures
may arise: the consequences of such a coupling are the subject of this paper.

The interplay between disorder and exchange for QHFMs
has been examined previously from 
several different viewpoints.
Fogler and Shklovskii, building on earlier 
discussions, have developed
a mean field treatment in the spirit of Stoner theory,
finding for odd integer $\nu$ in the absence of Zeeman coupling
a transition between ferromagnetic and
paramagnetic ground states with increasing disorder strength \cite{shklovskii}.
They suggested that this transition should be apparent in transport 
measurements, in which the ferromagnetic phase is characterised by 
spin-resolved Shubnikov-de Haas oscillations and the paramagnet by
spin-unresolved oscillations. Experimentally, a  transition
of this kind is observed with decreasing magnetic field 
strength \cite{expt}, and its sharpness suggests that its
origin is indeed cooperative.
Within such an approach, proposed for higher
Landau levels where skyrmions are normally not stable, local moments 
are all collinear in the ferromagnet and vanish in the paramagnet.
By contrast, near $\nu=1$, an alternative is that
a QHFM may respond to 
disorder mainly via the direction rather than the magnitude
of its local magnetisation. Some indications that this can happen 
come from calculations for
the fully-polarised ferromagnet at weak disorder. Here,
a reduction in spin stiffness with increasing disorder strength
has been interpreted by Green \cite{green1} as
a precursor of a non-collinear phase.
Moreover, even weak disorder may 
nucleate a dilute glass of skyrmions (over minima in the electrostatic
potential) and antiskyrmions (over maxima), as discussed by 
Nederveen and Nazarov \cite{nazarov}. In addition, at intermediate
disorder strength both reduced and non-collinear local moments
emerge from a numerical solution of Hartree-Fock theory for a model
with Coulomb interactions and spatially uncorrelated disorder, by
Sinova, MacDonald and Girvin \cite{sinova}. More generally, the
relative importance for dirty QHFMs 
of local moment reduction versus the formation of
spin textures will depend on the nature of disorder. In
the following we focus on textures, favoured by a smoothly varying
impurity potential.

To this end, consider a quantum Hall system with $\nu$ close to unity, 
impurity potential $V({\bf r})$, electron density $\rho({\bf r})$, and
electron-electron interaction energy U({\bf r}).
As a first step, treat screening using Thomas-Fermi theory,
omitting exchange interactions and Zeeman energy. Within this approximation,
developed by Efros \cite{efros} for the comparable problem in spin-polarised
Landau levels when $\nu$ lies near half-integer values, the ground-state charge
density at weak disorder is determined by the 
condition that the 
Hartree potential should everywhere match the chemical potential: 
$
\mu=V({\bf r})+ 
\int  U({\bf r}-{\bf r}') \rho({\bf r}')   d^2{\bf r}'\,.
$
We are concerned with circumstances in which the resulting
density varies smoothly on the scale of the
magnetic length, $l_{\rm B}$, and has fluctuations from $\nu=1$,
$\delta \rho({\bf r}) \equiv \rho({\bf r})-(2\pi l_{\rm B}^2)^{-1}$,
which are small: $|\delta \rho({\bf r})| \ll \rho({\bf r})$.
Restoring exchange interactions
under these conditions results locally in maximal ferromagnetic polarisation
of electron spins, with a direction that may vary in space.
Denoting this direction by the three-component unit vector 
$\vec{S}({\bf r})$, its spatial fluctuations are linked to electron 
density via \cite{sondhi,rajaraman}
\begin{equation}
\delta\rho({\bf r})= (8\pi)^{-1} \epsilon_{ij} \epsilon_{\alpha \beta \gamma}
S^{\alpha}\partial_i S^{\beta} \partial_j S^{\gamma}\,.
\label{rho}
\end{equation}
An exchange energy is associated with such variations: combining
exchange (with interaction constant $J$), impurity and Hartree contributions
to the total energy, and choosing for simplicity 
a short-range interaction
$U({\bf r}-{\bf r}')= U_0\delta({\bf r}-{\bf r}')$, we take as our description
of a dirty quantum Hall ferromagnet the configurational energy
\begin{equation}
{\cal H}=\int\left( \frac{J}{2} |\nabla \vec{S}({\bf r})|^2  +
V({\bf r}) \delta \rho({\bf r})  + 
\frac{U_0}{2} [ \delta \rho({\bf r})]^2 \right) d^2{\bf r} \,.
\label{model}
\end{equation}
Doing so, we neglect quantum fluctuations of $\vec{S}({\bf r})$, as is 
justified in the semiclassical limit, 
$|\nabla \vec{S}({\bf r})| \ll l_{\rm B}^{-1}$,
to which we are already restricted. 
We choose for simplicity $V({\bf r})$ Gaussian distributed with
amplitude $\Delta$ and correlation length $\lambda$.
(More realistic choices for 
$U({\bf r})$ and for the distribution of $V({\bf r})$
will be considered elsewhere \cite{long}.)

Our aim in the following is to understand the zero-temperature phase
diagram of the model defined by Eq.\,(\ref{model}), and to characterise its 
ground states via their response functions and excitations. As an
example of a disordered electron system, it is unusual in
that there is an exchange gap for single-particle excitations,
even if the ground-state spin configuration 
$\vec{S}({\bf r})$ is not ferromagnetically ordered,
so that the only low-energy excitations involve collective spin modes. 
As an example of a ferromagnet with quenched disorder, 
the system is also unusual
in several ways. First, the coupling to disorder leaves spin-rotation
symmetry intact but breaks time-reversal symmetry, in contrast to
random magnetic fields, which break both symmetries, and to random exchange
interactions, which leave both symmetries intact. Second, because of the
same coupling, the spin system responds to applied electric fields: 
we calculate the wavevector-dependent dielectric susceptibility, comparing
with behaviour found in more conventional disordered
electron systems. Third, the link between spin and charge
also endows spinwaves with an electric dipole moment: we calculate the
spinwave contribution to the optical conductivity in disordered spin states, 
complementing Green's results 
\cite{green1} for the polarised ferromagnet with impurities.

We start with a simple discussion of the phase diagram. The model
is characterised by two energy scales, $J$ and $\Delta$, and two
length scales, $\lambda$ and $L_{\rm H} \equiv (U_0/J)^{1/2}$.
The last of these, which we call the Hartree length, plays an
important role in what follows. Its significance can be made
clear by comparing, for a skyrmion of fixed
shape and radius $R$ in a clean system, the contributions to
total energy from exchange and from Hartree interactions,
of order $J$ and $U_0/R^2$ respectively: 
exchange dominates on lengthscales large compared with $L_{\rm H}$,
while Hartree interactions dominate at smaller distances.
We use the limit $L_{\rm H} \gg \lambda$ as a source of simplifications
in analytical estimates, but take $L_{\rm H} \sim \lambda$ in
numerical simulations.

Examine first the ground state for $\langle \delta \rho \rangle = 0$
as a function of $\Delta/J$. 
At weak disorder
($\Delta \lesssim J$) it can be pictured as a dilute glass
of skyrmions and antiskyrmions, discussed in Ref.\,\cite{nazarov}.
This reflects the existence of a threshold \cite{green1,nazarov,sinova},
defined by $|V({\bf r})|=4\pi J$: in most parts of the system
(roughly, those where $|V({\bf r})|$ is below threshold) ferromagnetic order
is essentially unaffected by the impurity potential 
\cite{footnote}. More precisely, for any $\vec{S}({\bf r})$
one has $|\nabla \vec{S}({\bf r})|^2 \geq 8\pi |\delta \rho({\bf r})|$ and hence
${\cal H} \geq  \int [4\pi J |\delta \rho({\bf r})| + 
V({\bf r})\delta \rho({\bf r})] d^2{\bf r}$, which for $V({\bf r})$ 
below threshold everywhere implies that the ground state is 
the perfectly aligned
ferromagnet with $\delta \rho({\bf r})=0$. At stronger disorder 
($\Delta \gg J$), by contrast, screening is almost perfect at 
short distances and
$\delta \rho({\bf r}) \simeq - V({\bf r})/U_0$.
Corrections to such screening arise at and beyond the scale $L_{\rm H}$, 
where exchange is important. We can summarise the effect of exchange 
by dividing the
system into regions of area $L_{\rm H}^2$, 
finding for each such area the integral 
$Q\equiv -\int V({\bf r})/U_0\, d^2{\bf r}$, and adjusting the total
screening charge within every region to the integer value closest to $Q$.
We argue that these integers are predominantly zero in a ferromagnetic
phase, and predominantly non-zero in a phase without ferromagnetic order.
To see this, consider
a well-ordered ferromagnetic phase, in which ${S}({\bf r})$ as a function
of ${\bf r}$
has small fluctuations 
around a global direction of magnetisation. In this case,
the net topological charge in each region has magnitude much less than one.
Conversely, in a phase without such order, unit topological charge accumulates
in a region of linear size given by the ferromagnetic correlation length.
The phase boundary for the ferromagnet is therefore located by setting
$\langle Q^2 \rangle^{1/2}\sim 1$. Since each area of size $\lambda^2$
contributes to $Q$ a charge of magnitude $\lambda^2\Delta/U_0$ with random
sign, $\langle Q^2 \rangle^{1/2}\sim \lambda L_{\rm H}\Delta/U_0$ yielding
$\Delta_c \sim U_0/(\lambda L_{\rm H}) = J(L_{\rm H}/\lambda)$. For 
$\Delta > \Delta_c$ the ground state has no ferromagnetic order; because
within a classical zero-temperature description spins are frozen,
we identify this phase as a spin glass. 

\begin{figure}
\epsfxsize=2.375in \epsfxsize=3.25in
\centerline{\epsffile{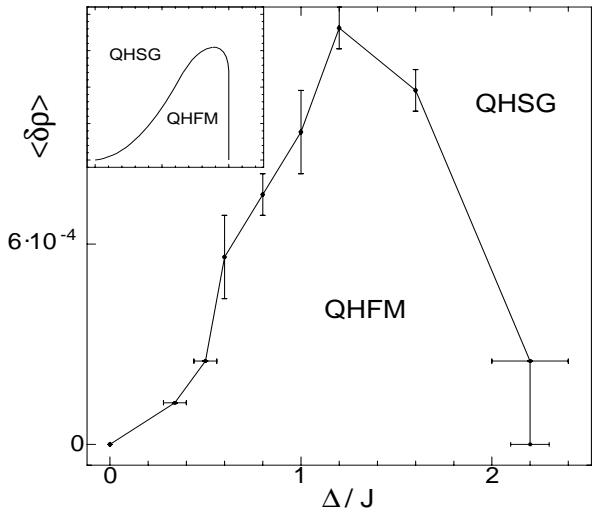}} \vspace{0.0in} \caption{
Phase diagram for $L_{\rm H}= \sqrt{2}\lambda$, obtained from
simulations. Inset: phase diagram for $L_{\rm H}\gg \lambda$, 
obtained from analytical arguments described 
in the text.} 
\label{fig1}
\end{figure}

Consider next the ground state at fixed $\Delta < \Delta_c$, as
a function of $\langle \delta \rho \rangle$. Charge is introduced into 
the system for $\langle \delta \rho \rangle > 0$ as skyrmions of 
size $R$. In the presence of Hartree energy alone, $R$ is divergent, but
fluctuations of an impurity potential establish an optimal size,
by generating random potential wells. For 
$R \gg \lambda$ these have an average depth $\Delta(\lambda/R)$
(the case $R \lesssim \lambda$ is treated in Ref.\,\cite{nazarov}). 
The contributions to the skyrmion
energy from these two sources are of order $U_0/R^2$ 
and $-\Delta\lambda/R$, respectively; minimising their sum, we find 
$R\sim U_0/(\Delta \lambda) = L_{\rm H}(\Delta_c/\Delta)$.
(Note that, since $R>L_{\rm H}$ for $\Delta < \Delta_c$, screening
as discussed in the previous paragraph does not alter this argument.)
We expect ferromagnetic order to persist with increasing 
$\langle \delta \rho \rangle$ until such skyrmions overlap,
so the phase boundary lies at 
$\langle \delta \rho \rangle_c \sim L_{\rm H}^{-2}(\Delta/\Delta_c)^2$.
In this way we arrive at the schematic phase diagram shown in the inset to Fig.\,\ref{fig1}.

In order to test these arguments, we have studied ground states of 
a lattice model numerically. In outline, we replace Eq.\,(\ref{model}) with
a Heisenberg model on a square lattice which has nearest-neighbour 
ferromagnetic interactions of strength $J$ and a
topological charge defined on each elementary plaquette.
These charges contribute to the total energy by
a local Hartree interaction with coefficient $U_0$, and by
their interaction with plaquette potentials,
independently and uniformly distributed in $[-\Delta, \Delta]$.
Thus $\lambda=1$ in units of the lattice spacing.
We find low-energy states using a simulated annealing procedure 
similar to that of Ref.\cite{anneal},
and can identify states that are equivalent under a global rotation
by comparing total energies and also charge distributions.
Using system sizes up to $56^2$ spins and annealing with $10^6$
Monte Carlo steps per spin, for a given disorder realisation
in successive runs we repeatedly reach states from a small set:
we take lowest of these in energy to be the ground state.
In the ground state for each disorder realisation 
we calculate the site-averaged magnetisation,
$M=|\langle \vec{S}({\bf r})\rangle|$. We also determine: the
ground-state susceptibility $\chi$, from the response to a Zeeman field
applied in the direction of the magnetisation; the ground-state spin stiffenss
$\rho_{\rm S}$,
from the energy cost of long-wavelength twists imposed on the spin
configuration; and the wavevector-dependent compressibility,
from the linear response of $\delta \rho({\bf r})$ to a
periodic potential added to $V({\bf r})$.
Further details of our methods  
will be described elsewhere \cite{long}.

\begin{figure}
\epsfxsize=2.375in \epsfxsize=3.25in
\centerline{\epsffile{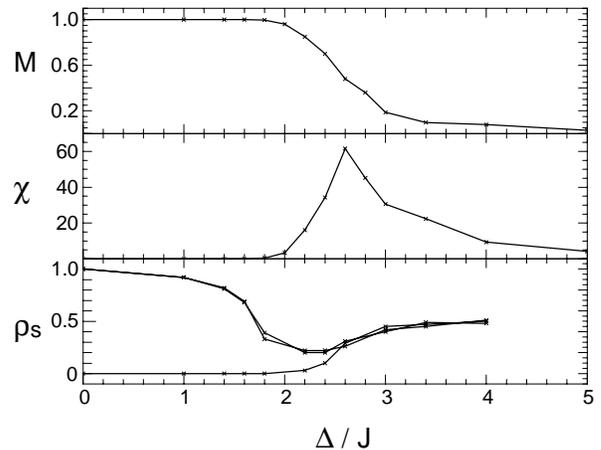}} \vspace{0.0in} \caption{Ground-state
properties as a function of disorder strength: magnetisation $M$,
susceptibility $\chi$, and eigenvalues $\rho_{\rm S}$ of the spin 
stiffness tensor.} 
\label{fig2}
\end{figure}

Representative results are shown in Fig.\,\ref{fig2}. Taking 
$\langle \delta \rho \rangle =0$ and $U_0/J=2$, 
we present $M$, $\chi$ and $\rho_{\rm S}$
as a function of disorder strength, $\Delta/J$, using $40^2$ spins
and an average over three disorder realisations.
Three regimes are evident from the behaviour of $M$. 
For $\Delta/J < 1$,
the ground state is a fully-polarised ferromagnet, because we use
a bounded disorder distribution. For $1 < \Delta/J \lesssim 2.5$, the
ground state is a partially polarised ferromagnet
(though with an almost saturated magnetisation for 
$1 < \Delta/J \lesssim 1.8$). And for 
$2.5 \lesssim \Delta/J$, the ground state is, we argue, a spin glass,
with a small non-zero $M$ arising as a finite-size effect.
In support of this interpretation, a large peak in $\chi$ 
indicates a phase transition at  $\Delta/J\approx 2.5$.
Moreover, the distinction between phases is illustrated by 
their spin stiffness, specified in full by a $3\times 3$
symmetric tensor. We plot the eigenvalues of this in Fig.\,\ref{fig2}.
For the fully-polarised ferromagnet, rotations about the magnetisation
direction do not alter the spin configuration, and so one eigenvalue is
zero while the other two are degenerate
For the partially-polarised
ferromagnet, all three eigenvalues are non-zero, with two remaining degenerate.
And at $2.5 \lesssim \Delta/J$, magnetic isotropy and finite
spin stiffness in the spin glass are illustrated by the fact that
all three eigenvalues are approximately degenerate and non-zero.
 
Repeating such calculations for $\langle \delta \rho \rangle >0$, we 
determine the phase diagram shown in the main panel of 
Fig.\,\ref{fig1}, which is 
qualitatively similar to that in the inset, even though the
simulations are for $(U_0/J\lambda^2)=2$,
while our theoretical arguments apply for $(U_0/J\lambda^2) \gg 1$.
The maximum density range over which disorder may stabilise the
ferromagnet, $|\langle \delta \rho| \lesssim 10^{-3}$ in units
of charge per plaquette, is strikingly narrow.

We now turn to the dielectric response of the partially polarised ferromagnet
and spin glass,
characterised at zero frequency by the wavevector-dependent dielectric
susceptibility $\chi_{\rm e}(q)$ or by the compressibility $\kappa(q)$, with 
$\kappa(q)=q^2\chi_{\rm e}(q)\epsilon_0/e^2$. To fix definitions, take
$V({\bf r}) \to V({\bf r}) + V_1 \cos({\bf q}\cdot {\bf r})$ in
Eq.\,(\ref{model}): the ground-state electron density changes according to
$\delta \rho({\bf r}) \to \delta \rho({\bf r})
+ \delta \rho_1({\bf r})$ and the disorder-averaged linear response
is $\langle \delta \rho_1({\bf r})\rangle = 
-\kappa(q)V_1\cos({\bf q}\cdot {\bf r})$. We can apply to $\kappa(q)$
the approach used in our discussion of the phase diagram: the Hartree
length $L_{\rm H}$ again plays an important role, within
our model problem in which Coulomb interactions are omitted. 
For $q\gg L_{\rm H}^{-1}$, exchange may be neglected and 
$\kappa(q)\simeq U_0^{-1}$. 
More generally, we anticipate the scaling form 
$\kappa(q) = U_0^{-1} f(L_{\rm H}q)$, with $f(x)$ constant
at large $x$.
For $q\ll L_{\rm H}^{-1}$ we expect that exchange dominates
and hence that $\kappa(q)$ should be independent of $U_0$ in this regime. 
In turn this implies for the scaling function $f(x) \propto x^2$
at small $x$, so that
$\kappa(q)\propto (qL_{\rm H})^2 U_0^{-1}\equiv Jq^2$ for $q\ll L_{\rm H}^{-1}$.
To summarise, the response is that of a metal ($\kappa(q)$ constant) at 
large wavevectors, and that of an insulator ($\chi_{\rm e}(q)$ constant) at
small wavevectors \cite{footnote2}. 

\begin{figure}
\epsfxsize=2.375in \epsfxsize=3.25in
\centerline{\epsffile{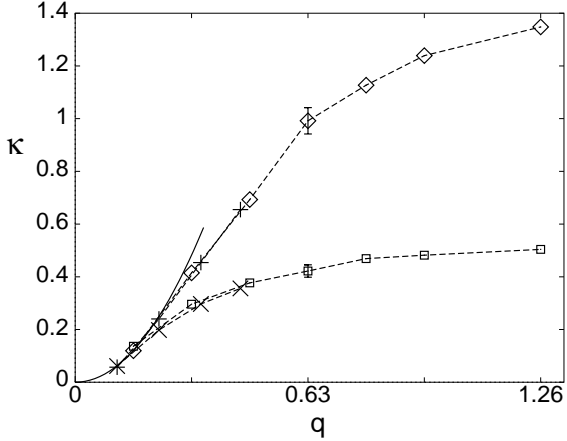}} \vspace{0.0in} \caption{Compressibility
$\kappa(q)$ as a function of wavevector $q$ for systems with $J=1$
and $U_0=1$ ($\Diamond$,$+$) or $U_0=2$ ($\Box$,$\times$);
full line: quadratic fit at small $q$.}
\label{fig4}
\end{figure}

We have tested these arguments using the simulation methods outlined above.
Results for $\kappa(q)$ are displayed in Fig.\,\ref{fig4}, where we
compare behaviour in systems 
with $J=1$ and $U_0=1$ or $U_0=2$, 
combining for each case data from lattices of size $40^2$ and $56^2$
in order to maximise
wavevector resolution. The 
distinction is clear between a $U_0$--independent
$\kappa(q)$, quadratic in $q$, at small $q$, and a 
$q$--independent $\kappa(q)$, varying roughly as $U_0^{-1}$, at large $q$.

Finally, we consider optical conductivity $\sigma(\omega)$ at frequency $\omega$.
A spin glass in the absence of Zeeman energy is expected 
\cite{halperinginzburg} to support three degenerate Goldstone modes.
Their dispersion is linear at small frequencies with speed
$c=(\rho_{\rm S}/\chi)^{1/2}$. In an QHFM they have an electric
dipole moment, which arises because they propagate in a non-collinear
ground state. Their contribution to $\sigma(\omega)$ is determined
by a product of the mean square dipole moment with the density of states.
We find \cite{long}
\begin{equation}
\sigma(\omega) \sim 
\frac{e^2}{h}\left(\frac{\omega \xi}{c}\right)^2 \frac{\hbar \omega}{J}
\end{equation}
where $\xi$ is the spin correlation length.
Variable-range hopping would mask this contribution to $\sigma(\omega)$
but should be absent if spins locally have maximal polarisation
throughout the sample.

In summary, we have shown how the competition between screening and exchange
determines the ground-state properties of QHFMs
in the presence of a smoothly-varying impurity potential.

We are grateful for discussions with N. R. Cooper and S. L. Sondhi.
The work was supported in part by the EPSRC under 
Grant GR/J78327 (JTC), and by the Royal Society (DKKL).

\end{document}